\documentstyle[prl,aps,psfig,twocolumn,floats]{revtex}
 
\begin{document}

\title{Pulling strings at finite temperature:\\
Another force-extension formula for the Worm-Like Chain Model}
\author{Henrik Flyvbjerg}

\address{Condensed Matter Physics and Chemistry Department,
         Ris\o{} National Laboratory, DK-4000 Roskilde, Denmark\\
         and\\
   The Niels Bohr Institute, Blegdamsvej 17, DK-2100 Copenhagen \O, Denmark} 
\date{August 1, 2000}
\maketitle

\begin{abstract}
The derivation of Marko and Siggia's interpolation formula 
for the force-extension relation of the Worm-Like Chain Model 
( C.~Bustamante, J.~F.~Marko, E.~D.~Siggia, and S.~Smith, 
   Science {\bf 265}, 1599 (1994);
J.~F.~Marko and E.~D.~Siggia, Macromolecules {\bf 28}, 8759 (1995)) 
is retraced.
Isotropy of space, resulting in rotational invariance of the free energy, 
is invoked together with analyticity.
A new interpolation formula results for the force-extension relationship.
It is as simple as the old one,
but twice as close to the exact force-extension relationship.
Application of the same reasoning to the second-order perturbative result
obtained at strong force (J.~D.~Moroz and P.~Nelson, 
Proc.~Natl.~Acad.~Sci.\ USA {\bf 94}, 14418 (1997)) 
results in yet a new interpolation formula,
good to 1\%\ at all forces.

\end{abstract}

\narrowtext
\paragraph*{Introduction.}

The worm-like chain (WLC) model 
\cite{aka,Kratky_Porod_49,Fixman_Kovac_73,Kovac_Crabb_82,Doi_Edwards_1986}
is the quintessential model of entropic elasticity 
from a flexible, but unstretchable fiber, string, or thread.
The model is conceptually simple, mathematically minimalist,
and widely used to interpret experiments 
that involve pulling at strings at finite temperature.
Thus recent single-molecule experiments in biological physics 
cause new interest in this old model from polymer physics.
It was successfully employed to model 
the experimental force-extension relationship of double-stranded DNA
\cite{Bustamante_et_al_94,Marko_Siggia_95,Wang_et_al_97} some years ago.
Also, the force-extension relations for the giant muscle protein titin 
\cite{Rief_et_al_97_titin,Rief_Fernandez_Gaub_98},  
the polysaccharide dextran 
\cite{Rief_Fernandez_Gaub_98,Rief_et_al_97_dextran},
and single molecules of xanthan \cite{Li_Rief_Oesterhelt_Gaub_99}
were explained with the WLC model.
A refined analysis of DNA's force-extension relationship
in terms of the WLC model was recently presented in \cite{Bouchiat_et_al_99}.
The relaxation dynamics of extended DNA molecules,
measured with millisecond resolution and femtonewton force spectroscopy,
was interpreted using the WLC model in \cite{Meiners_Quake_2000}.
In \cite{Baumann_et_al_00}, the WLC model was used to
interpret stretching of single collapsed DNA molecules.
Furthermore, the WLC model was extended with {\em stretch} 
\cite{Wang_et_al_97,Bouchiat_et_al_99}
and {\em twist} \cite[and references therein]
{Marko_Siggia_94_June,Moroz_Nelson_97,Moroz_Nelson_98,Bouchiat_Mezard_98}
to model also these property of double stranded DNA.

\paragraph*{The WLC model.}

The  WLC-models describes a string of vanishing cross section,
unstretchable, but flexible.
As it cannot stretch, the string can only bend,
and it resists even that with a force (per unit length) 
proportional to its curvature.
The constant of proportionality, $A$, 
is called the {\em bending modules},
and has dimension energy per unit length of string, 
per unit curvature squared.
Thus the bending energy of the string is
\begin{equation}
\label{eq:energy}
E[\vec{t}] = 
  \frac{A}{2} \int_0^{L_0} ds \left( \frac{d\vec{t}}{ds}(s) \right)^2
  \enspace,
\end{equation}
where $L_0$ is the length of the string;
$s$ parametrizes points on the string by their
distance from one end, as measured along the string,
$s\in [0,L_0]$; 
$\vec{t}(s)$ is the tangent vector to the string at its point at $s$.
This energy is evidently minimal for a straight string,
since a straight string has constant tangent vector.

When the string is submerged in a heat bath,
a bending energy of order $k_{\rm B}T$ is available 
to each of its degrees of freedom. 
The string consequently bends in a random manner
obeying Boltzmann statistics,
and an attempt to pull apart its ends
is resisted with a force $\vec{F}$ 
which depends on the string's end-to-end separation $\vec{R}$,
$\vec{F} = \vec{F}(\vec{R})$.
Because of the isotropy of space and the rotational invariance of
the bending energy in Eq.~(\ref{eq:energy}), 
$\vec{F}(\vec{R})$ is anti-parallel to $\vec{R}$,
and $|\vec{F}|$ depends only on $R\equiv|\vec{R}|$.
Because the string is unstretchable, the largest possible
end-to-end separation is $R = L_0$.
So a natural dimensionless measure of the end-to-end separation is
$r \equiv R/L_0 \in [0,1]$.

Despite its simplicity, 
the WLC model cannot be solved analytically in general.
When its ends are left free, 
a small calculation based on the Boltzmann
weight factor $\exp(-E[\vec{t}]/k_{\rm B}T)$,
results in the correlation function
$\langle \vec{t}(s_1) \cdot \vec{t}(s_2) \rangle =
\exp(-|s_1-s_2|/L_{\rm p})$,
where the {\em persistence length\/} $L_{\rm p}$ is inversely proportional
to the temperature: $L_{\rm p}=A/k_{\rm B}T$
\cite{Doi_Edwards_1986}.
In the limit $L_{\rm p}/L_0\rightarrow \infty$
where the string is 
much longer than its persistence length,
$L_{\rm p} F(r)/k_{\rm B}T$ is a dimensionless function
of the dimensionless variable $r$, only.
But not even in this convenient limit
is an exact analytical solution possible.
A numerical solution for the force-extension relationship
is not difficult to obtain
\cite{Marko_Siggia_95,Bouchiat_et_al_99}, 
and is given in a useful form in \cite{Bouchiat_et_al_99,formula}.

\paragraph*{The Marko-Siggia interpolation formula.}

It is sometimes convenient, however, to have a simple analytical expression
for the force-extension relationship,
even if only an approximate one.
Marko and Siggia presented such a relationship in 
\cite{Bustamante_et_al_94,Marko_Siggia_95}:
\begin{equation}
\label{eq:FMS}
\frac{L_{\rm p} F_{\rm MS}(r)}{k_{\rm B}T} = 
\frac{1}{4(1-r)^2} -\frac{1}{4} + r \enspace.
\end{equation}
This formula was derived
by calculating the force-extension relationship analytically
to leading order in the limit of very large force, 
where the string is nearly fully stretched, $r \approx 1$
\cite{Marko_Siggia_95}.  That yielded
\begin{equation}
\label{eq:Fone}
\frac{L_{\rm p} F_{\rm exact}(r)}{k_{\rm B}T} = 
\frac{1}{4(1-r)^2} + \mbox{unknown subdominant terms.} 
\end{equation}
An attempt to use the known part of this result 
for all forces/all $r\in [0,1]$
fails at small forces/small $r$, 
where a calculation \cite{Marko_Siggia_95} shows that
\begin{equation}
\label{eq:Fzero}
\frac{L_{\rm p} F_{\rm exact}(r)}{k_{\rm B}T} \sim
\frac{3}{2} r \mbox{~~for~~} r \sim 0
\enspace.
\end{equation}
The explicit term in Eq.~(\ref{eq:Fone}) does not satisfy Eq.~(\ref{eq:Fzero}).
The last two terms on the right-hand-side of Eq.~(\ref{eq:FMS})
were added to ensure Eq.~(\ref{eq:Fzero}) is satisfied.  
We note that this procedure 
does not compromise the validity of the result,
Eq.~(\ref{eq:FMS}), at $r \rightarrow 1$, 
because the two terms added remain finite in that limit.
The result is accurate to 17\%\ when at its worst---see Fig.~1---%
and is by construction asymptotically correct for $r\rightarrow 0$ and
for $r\rightarrow 1$.

\paragraph*{Another interpolation formula.}

Now consider the exact force-extension relationship.
Suppose we could calculate the string's free energy, ${\cal F}$, 
analytically at given temperature and end-to-end separation $\vec{R}$.
Then we could calculate its force-extension relationship as
\begin{equation} \label{deriv}
\vec{F}(\vec{R}) = -\frac{\partial{\cal F}}{\partial \vec{R}} \enspace.
\end{equation}
We cannot do this, but we know that the free energy
is independent of the {\em direction\/} of $\vec{R}$.
It depends only on $R\equiv |\vec{R}|$.
Furthermore, we expect the free energy to be an analytical function of 
$\vec{R}$ for $|\vec{R}| < L_0$, 
hence analytical in $\vec{R}= \vec{0}$. 
Arguments for analyticity may be given \cite{analyticity},
or one may regard analyticity as a conjecture or postulate.
Or one may simply disregard the issue;
the interpolation formulas given below 
have the precision demonstrated in Fig.~1
no matter how we arrive at the formulas.
Faulty logic works fine here. 
 
Analyticity of ${\cal F}(\vec{R})$ in $\vec{R}= \vec{0}$
implies that ${\cal F}(\vec{R})$ has a Taylor series expansion in 
powers of $\vec{R}$'s components.
Rotational invariance consequently implies that 
$\vec{R}$'s components only occur in the combination $\vec{R}^2$
in this Taylor series.
Consequently,
${\cal F}$ is an analytical function of $\vec{R}^2$, 
${\cal F} = {\cal F}(\vec{R}^2)$. 
Thus ${\cal F}$ is an {\em even\/} analytical function of $\vec{R}$.
From Eq.~(\ref{deriv}) then follows that $\vec{F}(\vec{R})$
must be an {\em odd\/} analytical function of $\vec{R}$.

Marko and Siggia's interpolation formula, Eq.~(\ref{eq:FMS}),
is not an odd analytical function of $r$.
But if we retrace its derivation from Eq.~(\ref{eq:Fone}),
we note that this expression already contains what it takes 
to mend it:
we extrapolate Eq.~(\ref{eq:Fone}) from the limit $r \rightarrow 1$ 
to lower values of $r$ 
in an odd  manner by realizing that
the factor 1 in the numerator really is $r$,
while the factor 4 in the denominator is  $(1+r)^2$.
Thus we arrive at a new analytical interpolation 
formula for the force-extension relationship
of the WLC model:
\begin{equation}
\label{eq:Fnew}
\frac{L_{\rm p} F_{\rm 8\%}(r)}{k_{\rm B}T} = 
\frac{r}{(1-r^2)^2}  + \frac{1}{2}r \enspace.
\end{equation}
The last term on the right-hand-side has been added to ensure
that Eq.~(\ref{eq:Fzero}) is satisfied,
entirely in the spirit of Marko and Siggia's derivation
of their formula.

$F_{\rm 8\%}(r)$ should be a better approximation than $F_{\rm MS}(r)$,
because it respects rotational symmetry 
and correctly captures all odd terms in the Taylor series for
the exact result.
They vanish in the exact result, and they vanish in $F_{\rm 8\%}(r)$.
Figure~1 illustrates this improvement:  
$F_{\rm 8\%}(r)$ is much closer to the exact result than 
$F_{\rm MS}(r)$ is. 
Its differs less than 8\%, at most,
while $F_{\rm MS}(r)$ differs up to 17\%. 
At low values of $r$,  $F_{\rm 8\%}(r)$
represents an improvement by much more than a factor two. 
For larger values of $r$, the improvement is approximately
a factor two.  
Note that $F_{\rm 8\%}$ achieves this doubled precision
with half as much ``patching'' at $r\sim 0$: 
only one correcting term was added.

\begin{figure}[t] 
  \begin{center}
    \leavevmode    
    \psfig{figure=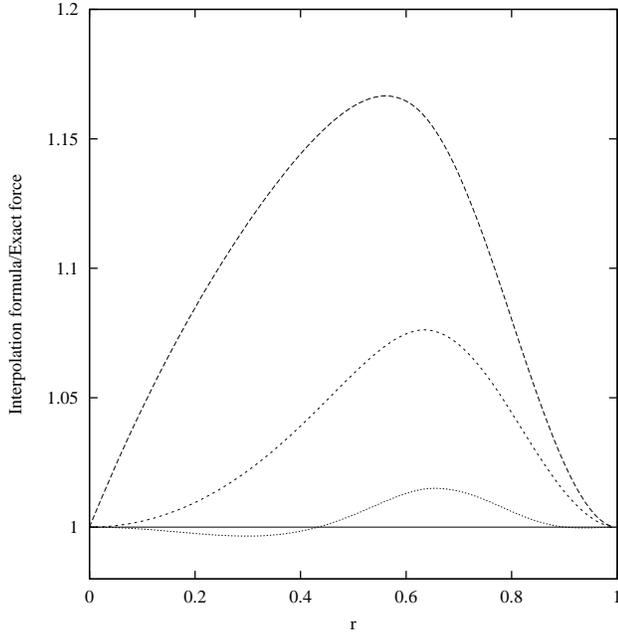,width=8.65cm}
    \caption[]{$F_{\rm MS}/F_{\rm exact}$ (upper curve),
     $F_{\rm 8\%}/F_{\rm exact}$ (middle curve),
    and $F_{\rm 1\%}/F_{\rm exact}$ (bottom curve) plotted against $r$.
    The maximum of $F_{\rm MS}/F_{\rm exact}$ is 1.17, 
    and occurs at $r=0.56$.
    The maximum of $F_{\rm 8\%}/F_{\rm exact}$ is 1.08, 
    and occurs at $r=0.64$.
    The maximum of $F_{\rm 1\%}/F_{\rm exact}$ is 1.015, 
    and occurs at $r=0.66$.
    The ratios plotted here are 
    identical to the inverse of the effective persistence length suggested in
     \cite[Eq.~(14)]{Bouchiat_et_al_99} for this kind of comparisons.}
  \end{center}
\end{figure}

\paragraph*{Yet another interpolation formula.}

Moroz and Nelson have calculated the first correction term to 
Eq.~(\ref{eq:Fone}) \cite{Moroz_Nelson_97,Moroz_Nelson_98,note}:
\begin{equation}
  \label{eq:FMN}
\frac{L_{\rm p} F_{\rm MN}(r)}{k_{\rm B}T} = 
\frac{1}{4(1-r)^2} + \frac{1}{32} + {\cal O}\left(1-r\right) \enspace.
\end{equation}
This result also is not an odd function of $r$.
But again this is easily remedied. 
The first term on the right-hand-side was treated above,
and the next two terms are treated similarly, yielding
\begin{equation}
  \label{eq:FMNodd}
\frac{L_{\rm p} F(r)}{k_{\rm B}T} = 
\frac{r}{(1-r^2)^2} + \frac{r}{32} + {\cal O}\left(r(1-r^2)\right) \enspace.
\end{equation}
This last expression is an odd analytical function of $r$,
but it does not satisfy Eq.~(\ref{eq:Fzero}),
and we cannot mend that simply by adding terms
which remain finite for $r=1$,
as Marko and Siggia did.
But we can proceed entirely in their spirit,
and add terms of same order as the neglected terms,
in the present case ${\cal O}\left(r(1-r^2)\right)$.
Doing that, we arrive at
\begin{equation}
  \label{eq:Fonepct}
\frac{L_{\rm p} F_{\rm 1\%}(r)}{k_{\rm B}T} = 
\frac{r}{(1-r^2)^2} + \frac{r}{32} + \frac{15}{32}r(1-r^2) \enspace.
\end{equation}
Figure~1 shows that $F_{\rm 1\%}(r)$ reproduces $F_{\rm exact}$ 
to within 1.5\%.

\paragraph*{Discussion.}

It is clear from the procedure we have used
that one may continue it systematically by calculating
more terms in the two series for $F_{\rm exact}(r)$'s
asymptotic behavior at $r=0$ and $r=1$, respectively.
This exact asymptotic information
can then be incorporated in an increasingly complex result,
by including an increasing number of terms
of the general form $r^{2n+1}(1-r^2)^m$,
with suitable coefficients and exponents $n$ and $m$.
In view of the accuracy 
already achieved with Eq.~(\ref{eq:Fonepct}),
this is hardly worthwhile for most purposes.

For one purpose, however, it looks promising:
as a way to present a high-precision analytical 
interpolation formula
meant for numerical evaluation \cite{Flyvbjerg}.
At weak and strong force (small and large end-to-end separations)
where numerical methods typically fail unless
special care is taken, such a result
is exact to a chosen order in perturbation
theory.
And chosen properly, this order renders the result
uniformly good to a desired precision,
for all forces/end-to-end separations.
Bouchiet et al.'s numerical interpolation formula 
\cite{Bouchiat_et_al_99,formula}
is of this nature, being exact to leading order at small and large
force, and {\em uniformly\/} good to 1\%\ \cite{note2}. 

The improved force-extension formulas presented here 
remain valid when the WLC model is extended 
to describe a somewhat stretchable string 
as done in \cite{Wang_et_al_97}.
This because the extension 
consists in allowing $L_0$ to stretch in a simple manner
depending on $R$, while $F(r)= F(R/L_0)$ is left unchanged.

It may be of interest to apply the approach used here
to the WLC model extended with twist.
This model is analyzed with strong-force perturbation theory
in \cite{Moroz_Nelson_97,Moroz_Nelson_98,Bouchiat_Mezard_98}.
So the results obtained there might have their
range of validity for a given precision
extended down to lower force. 

\paragraph*{Acknowledgments.}

This work was prompted by work with the Optical Tweezer Group 
at the Niels Bohr Institute and by
{\em Workshop on Models of Biological Motion\/} at 
Collegium Budapest, June 19--22, 2000.
Thanks to T.~Vicsek for the invitation, 
and to F.~J\"ulicher, J.~Prost, and M.~D.~Wang for discussions.


\begin{thebibliography}{12}
\bibitem{note}
  Actually, what Moroz and Nelson calculated perturbatively and give
  in  \cite{Moroz_Nelson_97,Moroz_Nelson_98} is $r$
  as a series in $1/K$, 
  where $K\equiv\sqrt{L_{\rm p}F_{\rm exact}/k_{\rm B}T}$:
  $r=1-1/(2K)-1/(128K^3) + {\cal O}(1/K^4)$.
  They did this for a version of the WLC model extended to
  describe twist, as well.
  Ignoring twist and inverting the series, one arrives at 
  Eq.~(\ref{eq:FMN}) above.
\bibitem{note2} 
   The formula, Eq.~(\ref{eq:Fnum}), is much more precise than 1\%\
   in the range where the authors actually used it!
\end{thebibliography}
\end{document}